\documentclass[conference]{IEEEtran}

\usepackage{cite}
\usepackage{amsmath,amssymb,amsfonts}
\usepackage{algorithmic}
\usepackage{graphicx}
\usepackage{textcomp}
\usepackage{xcolor}
\def\BibTeX{{\rm B\kern-.05em{\sc i\kern-.025em b}\kern-.08em
    T\kern-.1667em\lower.7ex\hbox{E}\kern-.125emX}}
    
\usepackage{algorithmic}
\usepackage[linesnumbered,boxed,ruled,commentsnumbered]{algorithm2e}
\usepackage{newfloat}
\usepackage{listings}

\usepackage{amssymb}
\usepackage{amsmath}
\usepackage{bm}
\usepackage{booktabs}
\usepackage{multirow}

\usepackage{fancyhdr}
\makeatletter
\newcommand{\linebreakand}{%
  \end{@IEEEauthorhalign}
  \hfill\mbox{}\par
  \mbox{}\hfill\begin{@IEEEauthorhalign}
}
\makeatother

\begin{document}

\title{PearNet: A Pearson Correlation-based Graph Attention Network for Sleep Stage Recognition}

\author{
\IEEEauthorblockN{Jianchao Lu}
\IEEEauthorblockA{\textit{School of Computing} \\
\textit{Macquarie University}\\
Sydney, Australia \\
jianchao.lu@hdr.mq.edu.au}

\and

\IEEEauthorblockN{Yuzhe Tian}
\IEEEauthorblockA{\textit{School of Computing} \\
\textit{Macquarie University}\\
Sydney, Australia \\
yuzhe.tian@hdr.mq.edu.au}

\and

\IEEEauthorblockN{Shuang Wang}
\IEEEauthorblockA{\textit{School of Computing} \\
\textit{Macquarie University}\\
Sydney, Australia \\
shuang.wang@mq.edu.au}

\linebreakand

\IEEEauthorblockN{Michael Sheng}
\IEEEauthorblockA{\textit{School of Computing} \\
\textit{Macquarie University}\\
Sydney, Australia \\
michael.sheng@mq.edu.au}

\and

\IEEEauthorblockN{Xi Zheng}
\IEEEauthorblockA{\textit{School of Computing} \\
\textit{Macquarie University}\\
Sydney, Australia \\
james.zheng@mq.edu.au}

}

\maketitle
\thispagestyle{fancy}
\begin{abstract}
Sleep stage recognition is crucial for assessing sleep and diagnosing chronic diseases. Deep learning models, such as Convolutional Neural Networks and Recurrent Neural Networks, are trained using grid data as input, making them not capable of learning relationships in non-Euclidean spaces. Graph-based deep models have been developed to address this issue when investigating the external relationship of electrode signals across different brain regions. However, the models cannot solve problems related to the internal relationships between segments of electrode signals within a specific brain region. In this study, we propose a Pearson correlation-based graph attention network, called PearNet, as a solution to this problem. Graph nodes are generated based on the spatial-temporal features extracted by a hierarchical feature extraction method, and then the graph structure is learned adaptively to build node connections. Based on our experiments on the Sleep-EDF-20 and Sleep-EDF-78 datasets, PearNet performs better than the state-of-the-art baselines.
\end{abstract}


\section{Introduction}
The prevalence of sleep disorders and deprivation poses a significant burden on public health, affecting millions globally. Typically, sleep recordings are collected by sensors attached to various body parts~\cite{fallmann2019computational}. The recorded signals are called polysomnography (PSG), including electroencephalography (EEG), electrooculogram (EOG), and other physiological signals~\cite{roebuck2013review}. According to the American Academy of Sleep Medicine (AASM) sleep standard, recorded signals are divided into $30s$ sleep episodes named sleep epochs. Each of these can be divided into five different stages of sleep (W, N1, N2, N3, REM)~\cite{moser2009sleep}. It is a laborious and time-consuming manual labeling process since several sensors are needed to be attached to each subject over multiple nights to record PSGs. At present, sleep diagnostics and assessment still relies heavily on human expertise. Researchers have therefore attempted to develop automated systems to identify sleep stages.

Recent studies have demonstrated the ability of deep learning to automatically recognize sleep stages due to its powerful feature representation learning mechanism. In the PSG, for example, different sleep stages usually have different salient waves, and Convolutional Neural Networks (CNNs) and Recurrent Neural Networks (RNNs) are often used to learn appropriate salient wave representations. While these deep learning methods can achieve high accuracy when recognizing sleep stages, they are limited to using grid data as input for the models. When considering the external relationships of signals in different parts of the brain, however, due to the non-Euclidean nature of the brain region, graph structure data is better suited to represent the signal relationships between the electrodes attached to distinct brain areas than a grid input. Reference~\cite{jia2020graphsleepnet} developed a graph based deep model to address this challenge, in which each EEG channel collected from the electrodes behaves as a node of the graph, and the connections between the nodes are edges. However, their model is specially designed for external relationships of signals with multi-channels, and cannot handle problems related to the internal relationships between the segments of individual electrode signals in a specific brain region.

In consideration of the internal relationships between signal segments in a specific brain area, although the signal itself is a Euclidean data structure, its segments internal relations, like correlations and connections between signal segments, represent non-Euclidean geometry. In this study, we propose a Pearson correlation-based graph attention network, PearNet, to graphically model the internal relationships between the signal segments with different receptive fields in a specific brain area. We divide the thirty-second sleep EEG signal into a specific number of base segments of equal length (e.g., five-second slice) and extract salient waves from each segment. The spatial information of salient waves is captured from the base segments through a Spatial Convolutional Network, and then these spatial features are used as an input to extract the temporal features through a Temporal Convolutional Network.

To capture the temporal connection from various types of salient waves, a hierarchical feature extraction structure is designed where each level’s receptive field is gradually expanded in an attempt to obtain more diverse wave forms from neighboring segments. In PearNet, internal relationships between EEG signal segments are evaluated based on the spatial-temporal features derived from all levels. 
Since not every segment contributes to sleep stage recognition, correlations and connections between segments are established through a graph structure, based on which attention coefficients between the relevant segments are learned to represent their recognition contributions. In the graph structure, each segment corresponds to a node of the graph, and the connection between the segments corresponds to the edge of the graph.

The key contributions of the PearNet are summarized as follows:
\begin{itemize}
    \item We propose PearNet, a Pearson correlation-based graph attention neural network to model the internal spatial-temporal relationships between the segments of individual electrode signals in a specific brain 
    region\footnote{https://github.com/ITSEG-MQ/PearNet}. 
    %
    To the best of our knowledge, this is the first attempt to identify sleep stages from the pairwise intrinsic connections between signal segments derived by the graph structure of bio signals. 
   \item We design a node generation mechanism to produce spatial-temporal graph nodes, where 1D convolutions and a squeeze-and-excitation block are used to extract spatial information, while Temporal Convolutional Network (TCN) with dilated causal convolutions is employed to obtain temporal information. To learn the graph structure of nodes, we propose a novel adaptive graph structure learning method that is integrated with a Pearson correlation-based graph attention mechanism.
   \item We evaluate PearNet against the existing models. Experimental results demonstrate that the PearNet achieves state-of-the-art performance in sleep stage recognition. We also conduct an ablation and sensitivity study to better understand how our proposed solution works and point out some future directions. 
\end{itemize}

\section{Related Work}
\noindent{\bf Sleep Stage Recognition}.
Machine learning methods such as Support Vector Machine (SVM)~\cite{lajnef2015learning}, Random Forest (RF)~\cite{fraiwan2012automated} and K-means~\cite{gunecs2010efficient} were used 
in 
earlier research 
to recognize sleep stages. These methods, however, require the extraction of handcrafted features, which demands a great deal of prior knowledge. Due to this, many researchers have been turning to deep learning techniques for recognizing sleep stages. CNNs and RNNs have been widely used in sleep stage recognition. A CNN model with a joint classification-and-prediction framework~\cite{phan2018joint} was proposed to retrieve sleep power spectrum information hidden in EEG signals. A similar study was provided in~\cite{phan2019seqsleepnet} where the authors transformed raw EEG signals into power spectral images and used a hierarchical RNN to classify multiple epochs simultaneously. Moreover, a deep CNN was developed to improve the accuracy of sleep stage recognition by taking into account the transitional rules between each stage using a long short-term memory (LSTM)~\cite{sun2018deep}. Similarly, DeepSleepNet~\cite{supratak2017deepsleepnet} used Bi-directional Long Short-Term Memory (BiLSTM) to learn the transition rules among sleep stages. In addition, 
many studies developed attention mechanisms for recognizing sleep stages, such as
SleepEEGNet~\cite{mousavi2019sleepeegnet}, 
AttenSleepNet~\cite{eldele2021attention}, 
TS-TCCNet~\cite{eldele2021time}, and
SalientSleepNet~\cite{jia2021salientsleepnet}.

Though the models above can detect salient waves and learn transition rules from PSG signals, one major limitation remains: data input must be 
in grid form. In many cases, the data gathered on grids cannot be used to observe the intrinsic relationships of signals from an internal and external perspective because of their non-Euclidean nature. In this regard, the graph is an appropriate data structure.

\noindent{\bf Graph Neural Network}. 
In dealing with graph structure data, Graph Neural Networks with attention mechanisms have demonstrated enhanced performance. Existing GNNs, such as the Graph Attention Network (GAT)~\cite{velivckovic2017graph} used feed-forward neural networks to learn attention scores, while GraphSAGE~\cite{hamilton2017inductive} relied on cosine similarity to compute attention scores. However, these methods utilize a fixed graph structure, which is not optimal for sleep stage recognition. To address this issue, GraphSleepNet~\cite{jia2020graphsleepnet} adaptively learned the external relationships of EEG signals (graph structure) between different brain regions. Spatial-Temporal Graph Convolution was used in this model to extract spatial-temporal features of the nodes, whereas it cannot handle problems related to the internal relationships between the segments of individual electrode signals in a specific brain region.

\section{Preliminaries}
\begin{figure*}[!htp]
\centering
\includegraphics[width=0.83\textwidth]{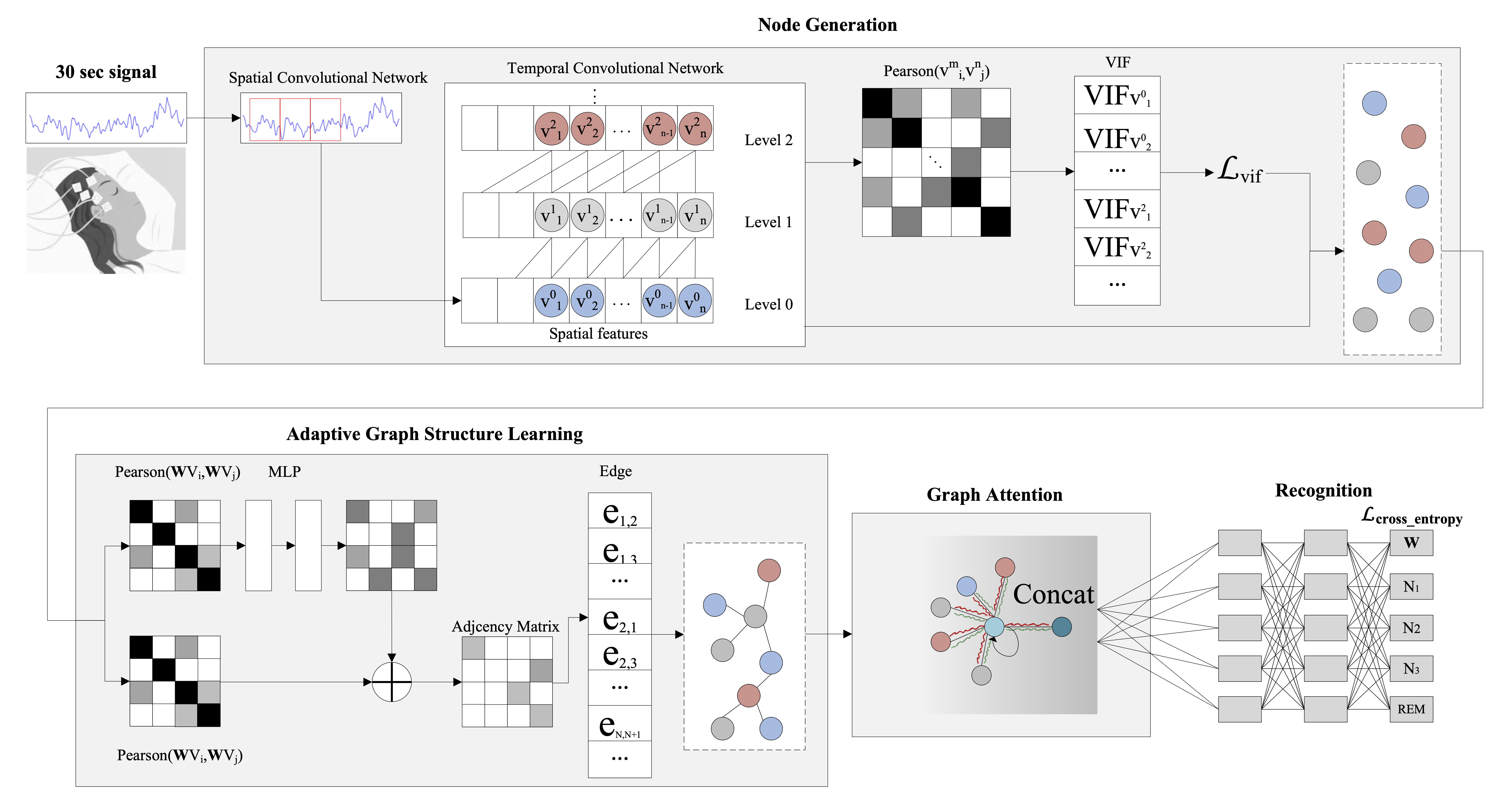}
\caption{The overall architecture of the proposed PearNet for sleep stage recognition.}
\label{fig:overall architecture}
\end{figure*}

In our study, a sleep stage recognition network is defined as an undirected graph $\cal{G = (V, E)}$, where ${\cal{V}}=\{v_{1}, v_{2}, ..., v_{n}\}$ denotes the set of $n$ nodes. Each node in the network represents a signal segment; ${\cal{E}}=\{e_{ij}\}_{v_{i},v_{j} \in {\cal{V}}}$ denotes the set of edges, indicating the connections between nodes. The raw signal sequence is defined as $S=\{s_{1}, s_{2}, ..., s_{L}\} \in \mathbb{R}^{L \times M}$, where $L$ is the number of samples, and $M$ is the number of data points of each sample.

The sleep staging problem is defined as: Given the raw signal sequence $S$, the signal graph nodes $\cal{V}$ are generated and the signal internal relationships (edges) $\cal{E}$ are learned to establish a graph $\cal{G}$ that maps $S$ into the corresponding sequence of sleep stages $\hat{Y}$, where $\hat{Y}$ = \{$\hat{y}_{1}$, $\hat{y}_{2}$, ..., $\hat{y}_{L}$\} and $\hat{y}_{i}$ is the recognition result of $s_{i}$. Following the AASM sleep standard, each $\hat{y}_{i} \in \{0, 1, 2, 3, 4\}$ matches with the five sleep stages W, N1, N2, N3, and REM, respectively.

\section{Method}
As illustrated in Figure~\ref{fig:overall architecture}, PearNet is a graph-based attention network, in which graph nodes are firstly derived from signal segments using spatial-temporal convolutions, followed by an adaptive graph structure learning method to construct the pairwise relationships between these nodes. After that, a Pearson correlation-based graph attention network is proposed to pay higher attention to the valuable nodes in the graph so as to recognize the sleep stages.

\subsection{Node Generation}
Node generation is the first key part of PearNet, which is used to extract spatial-temporal features from raw signals and generate graph nodes. As illustrated in Figure~\ref{fig:overall architecture}, the base segments of signals are predefined to capture the spatial information (spatial graph nodes) through a Spatial Convolutional Network, and then these spatial features are used as an input to extract the temporal features (temporal graph nodes) through a Temporal Convolutional Network. In general, the types of salient waves in different sleep stages vary. For example, the salient waves of stage N2 could be spindle waves or K-complex waves, while the salient waves of stage N3 are delta waves~\cite{olbrich2011multiple,jia2020graphsleepnet,yap2017oscillatory}. In order to capture the temporal connection from various types of salient waves, a hierarchical feature extraction structure is designed where each level’s receptive field is gradually expanded 
to obtain more diverse wave forms from neighboring segments (nodes). For example, Level 0 is the input layer which represents the spatial features extracted from the Spatial Convolutional Network, and the Level 1 layer denotes the temporal features extracted from the combination of each spatial feature segment (node) with its neighbours from the Level 0 layer so that the Level 1 layer contains more diverse wave forms. Similarly, the Level 2 layer refers to the temporal features extracted from combining each Level 1’s feature segment (node) with its neighbours' neighbours, and so on. In this study, our hierarchical feature extraction
is restricted to Level 2.

\subsubsection{Spatial Convolution:}
\begin{figure}[!tb]
\centering
\includegraphics[width=0.5\textwidth]{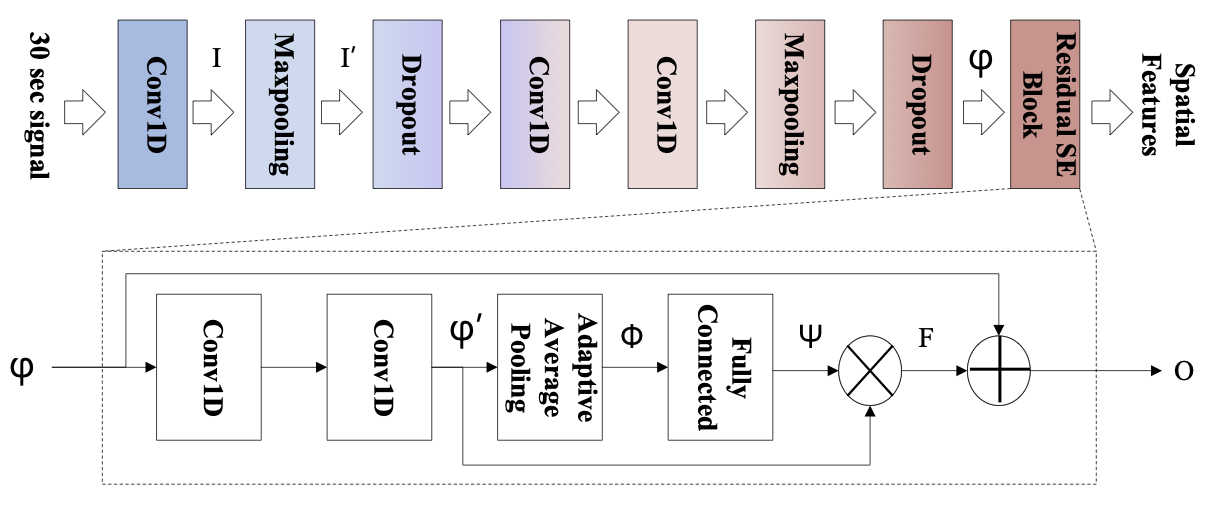}
\caption{Spatial Convolutional Network}
\label{fig:SCN}
\end{figure}
The Spatial Convolutional Network, as shown in Figure~\ref{fig:SCN}, starts from a 1D convolution layer with the input $X = \{x_{1}, x_{2}, ... , x_{M}\}, x_{i} \in  \mathbb{R}$, where $M$ is a hyperparameter that represents the length of a base segment. After that, a max pooling is used to down sample the feature map from $I = Conv1d(X), I \in \mathbb{R}^{C \times L}$ to $I' = Maxpooling(I), I' \in \mathbb{R}^{C' \times L'}$, where $C$ and $C'$ denote the 1D convolution output size before and after the max pooling, $L$ and $L'$ refer to a length of feature sequence before and after the max pooling. Following this, two 1D convolution layers and another max pooling are sequentially implemented to $I'$ such that $\varphi = Maxpooling\big(Conv1d\_2(Conv1d\_1(I'))\big), \varphi \in \mathbb{R}^{C'' \times L''}$, where $C''$ is the output size after two 1D convolution layers and max pooling, and $L''$ is the dimension of the feature output. Afterward, we utilize a residual Squeeze-and-Excitation (SE) block~\cite{hu2018squeeze} to recalibrate the features learned from previous convolutional layers, thus improving the performance. Given the feature map $\varphi$ learned from previous convolutional layers, we apply two 1D convolution layers to $\bm\varphi$ such that $\bm{\varphi'} = Conv2(Conv1(\varphi)), \varphi' \in \mathbb{R}^{N \times D}$, where $N$ is the output dimension, $D$ is the length of features, and $Conv1$ and $Conv2$ are two 1D convolution operations. A method of adaptive average pooling~\cite{gu2018stack} is then used to compress spatial information, thereby shrinking $\varphi'$ to $\phi = Avgpooling(\varphi'), \phi \in \mathbb{R}^{N \times D}$. Next, two fully connected (FC) layers are applied to make use of the aggregated information $\psi = \delta(W_{2}\sigma(W_{1}(\phi))), \psi \in \mathbb{R}^{N \times D}$, where $\delta$ and $\sigma$ refer to Sigmoid and Rectified Linear Unit (ReLU) activation functions respectively, and $W_{1}$ and $W_{2}$ represent two FC layers. The feature map $\varphi'$ is subsequently scaled by $\psi$ as follows:
\begin{equation}
    F = \bm{\varphi'} \otimes \psi \label{eq:scaled function} 
\end{equation}
where $\otimes$ represents a point-wise multiplication between $\varphi'$ and $\psi$. Lastly, a shortcut connection is applied to combine $\varphi$ learned from previous convolutional layers with the enhanced features $F$ derived from the residual SE block:
\begin{equation}
   O = \sigma(\bm\varphi + F) \label{eq:residual SE block}
\end{equation}
where $\sigma$ is a ReLU activation function. The output $O$ is the spatial features extracted from the basis segments.

\subsubsection{Temporal Convolution:} 
\begin{figure}[!t]
\centering
\includegraphics[width=0.5\textwidth]{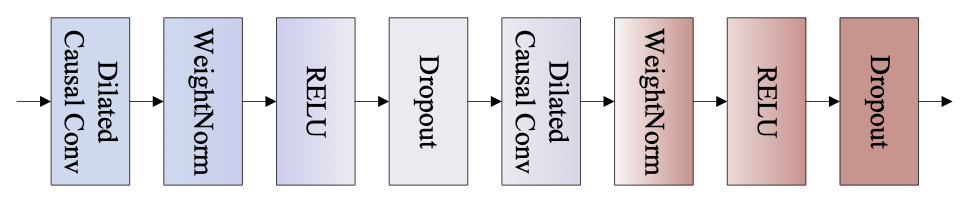}
\caption{Dilated Causal Convolution Block}
\label{fig:DCC}
\end{figure}

Based on the spatial features generated from the basis segments, TCN~\cite{bai2018empirical} is then used to extract temporal features. The main architecture is described as:
\begin{equation}
TCN = 1D~FCN + DCC
\end{equation}
where $1D~FCN$ is 1D fully-convolutional network, and $DCC$ (see Figure~\ref{fig:DCC}) is dilated causal convolution block with zero padding for keeping subsequent layers the same length as previous layers. 
Generally, the types of salient waves vary across sleep stages. In order to capture the temporal connection between different types of salient waves, a hierarchical level feature extraction structure is designed by using dilated convolutions to gradually enlarge the convolution's receptive field in an attempt to explore more types of salient waves. The dilated causal convolution $D$ in the $DDC$ on input $h$ is defined as~\cite{bai2018empirical}:
\begin{equation}
D(h)=(O*_{d}f)(h)=\sum_{i=0}^{k-1}f(i){\cdot}O_{h-d{\cdot}i}
\end{equation}
where input $O$ is the spatial features extracted from the base segments, $f:\{0,...,k-1\}{\rightarrow}\mathbb{R}$ is a convolutional filter, $k$ is the convolutional filter size, $d$ is the dilation factor which is used to determine the level of the hierarchical feature extraction, and $h$-$d{\cdot}i$ accounts for the direction of the past.

Using Spatial Convolution and Temporal Convolution, the node generation is summarized in Algorithm~\ref{alg:node generation}, where a spatial and temporal feature extraction process is described in lines 1 to 4. In line 6, the base segments (spatial features) and expanded segments (temporal features) corresponding to each level of the hierarchical feature extraction module are united as graph nodes.
\IncMargin{1em} 
\begin{algorithm}[!hbt]
    \SetAlgoNoLine 
    \SetKwData{Left}{left}\SetKwData{This}{this}\SetKwData{Up}{up} \SetKwFunction{Union}{Union}\SetKwFunction{FindCompress}{FindCompress} \SetKwInOut{Input}{Input}\SetKwInOut{Output}{Output}
    \Input{single channel EEG data $X = \{x_{1}, x_{2}, \cdots, x_{M}\}$, 
    the number of the base segments $S$,
    the level of feature extraction module $L$} 
    $h_{s} \leftarrow$ Spatial Convolution $\left(X, S\right)$\;
    $h_{t}\leftarrow \varnothing $\;
    \For { $l \in L$}{
        $h_{t}\leftarrow$   $h_{t}$ $\cup$ Temporal Convolution $\left(h_{s}, l\right)$\;
        }
    ${\cal{V}} \leftarrow h_{s} \cup h_{t}$ \;
    Return $\cal{V}$\;
    \caption{Node Generation}
    \label{alg:node generation}
\end{algorithm}
\DecMargin{1em}

\subsubsection{Node Generation Learning and VIF based Loss:}
Considering the dependency among these nodes generated from the hierarchical temporal convolution (three layers), it is likely that the generated nodes exhibit a multicollinearity property. Multicollinearity can lead to inaccurate or skewed results in the statistical models when determining how each variable is used to make a prediction~\cite{shen2020stable}. In order to address this issue, the weights of the spatial and temporal convolution networks in Node Generation are updated by minimizing the following loss function:
\begin{equation}
    \mathcal{L}_{vif} = \frac{1}{M}\sum_{i=1}^{M} smooth_{L1}(\delta(VIF_{i}) - \mathcal{T})
    \label{eq:loss function VIF}
\end{equation}
where $smooth_{L1}$ is Smooth L1-loss which combines the advantages of L1-loss (steady gradients) and L2-loss (less oscillations)~\cite{rui2020geometry}. $VIF_{i} \in (1, +\infty)$ is the variance inflation factor (VIF) for the $i$-th node generated from the spatial and temporal convolution networks, $\delta$ is a modified sigmoid function, $M$ is the total number of graph nodes, and $\mathcal{T}$ is the training target. We use VIF to determine the degree of multicollinearity~\cite{stine1995graphical}, defined as:
\begin{equation}
    VIF_{i} = \frac{1}{1-R_{i}^{2}} = \frac{M_{ii}}{|P|}
    \label{eq:VIF}
\end{equation} 
where $R_{i}^{2}$ is the $R^{2}$-value obtained by regressing the $i$-th node on the remaining nodes, $P$ is the Pearson correlation coefficient matrix of the graph nodes, $|P|$ is the determinant of the matrix, and $M_{ii}$ is the determinant of the Pearson correlation coefficient matrix without considering the value in the $i$-th raw and $i$-th column. The training target value $\mathcal{T}$ for VIF is 1, which indicates that the target node does not correlate with other nodes. Taking into account the range of $VIF_{i} \in (1, +\infty)$, a sigmoid function is used to normalize its value between 0 and 1 so that no rather large loss value will be presented. To facilitate the setting of the target value after the VIF passes through the sigmoid function, we let the function curve cross 0.5 at $\mathcal{T}=1$. Therefore, a modified sigmoid function is defined as follows:
\begin{equation}
    \delta(VIF_{i}) = \frac{1}{1+e^{(-VIF_{i}+1)}}
    \label{eq:modified sigmoid function}
\end{equation}  
After normalization through the modified sigmoid function, the new range of $VIF_{i} \in (0.5, 1)$, and the VIF training target $\mathcal{T}$ is changed from $1$ to $0.5$.

\subsection{Adaptive Graph Structure Learning}
The input to the adaptive graph structure learning is a set of nodes which contain spatial-temporal features generated from the node generation layer,  $\bm{h}$ = \{$\vec{h}_{1}$, $\vec{h}_{2}$, ..., $\vec{h}_{N}$\}, $\vec{h}_{i}$ $\in$ $\mathbb{R}^F$, where $N$ is the number of nodes, and $F$ is the dimension of features in each node. A learnable linear transformation, which is parameterized by a weight matrix, $\bm{W} \in \mathbb{R}^{F' \times F}$, is applied to input node features $\bm{h}$ in order to obtain sufficient expressive power to convert them into higher-level features. Self-attention is then performed on the transformed output with a shared attention mechanism $a : \mathbb{R}^{F'} \times \mathbb{R}^{F'} \to \mathbb{R}$ to compute attention coefficients,
\begin{equation} 
    e_{ij} = a(\bm{W}\vec{h}_{i}, \bm{W}\vec{h}_{j})
    \label{eq:attention coefficient}
\end{equation}
that indicate how node $j$ impacts node $i$~\cite{velivckovic2017graph}. 
As an alternative to graphs constructed manually or through prior knowledge (e.g., k-nearest neighbor graph~\cite{jiang2013graph}), we dynamically learn the graph structure. We define a non-negative function $A_{ij} = F(x_i, x_j)$ to represent the connection relationship between nodes $x_i$ and $x_j$ based on the attention coefficients ($e_{ij}$) of the nodes. $F(x_i, x_j)$ is implemented through double layer residual MLP. The learned graph structure (adjacency matrix) $A$ is defined as:
\begin{equation}
    A_{ij} = F(x_i, x_j) = \sigma\big(e_{ij} + (\bm{W_2}(\sigma(\bm{W_1}(e_{ij}))))\big)
    \label{eq:adjacency matrix}
\end{equation}
where $W_1$ and $W_2$ denote a double layer MLP, and $\sigma$ is a ReLU activation function that guarantees $A_{ij}$ is non-negative. The function $F(x_i, x_j)$ indicates that the graph structure is determined by the attention coefficients of nodes ($e_{ij}$) along with adjustment factors $\bm{W_2}(\sigma(\bm{W_1}(e_{ij})))$ for learning relationships under various sleep stages. We then construct the edges based on the adjacency matrix for all $A_{ij}$ that are greater than 0, and we remove the connections for those $A_{ij}$ with the value of 0.

\subsection{Graph Attention and Recognition}
For easy comparison of coefficients across different nodes, we use the softmax function to normalize them across all choices of $j$:
\begin{equation} 
    \alpha_{ij} = softmax(e_{ij}) = \frac{exp(e_{ij})}{\sum_{k \in {\cal{N}}_{i}}exp(e_{ik})}
    \label{eq:normalized coefficient}
\end{equation}
We employ a Pearson correlation coefficient-based attention mechanism to $a$ in Equation~\eqref{eq:attention coefficient}. While negative correlations play the same important role as positive correlations, their negative value can also cause the central nodes' influence to wane when aggregating the nodes. As a result, in Equation~\eqref{eq:normalized coefficient}, we use absolute Pearson correlation to compute the attention coefficients:
\begin{equation} 
    \alpha_{ij} = \frac{exp(\beta \times|Pearson(\bm{W}\vec{h}_{i}, \bm{W}\vec{h}_{j})|)}{\sum_{k \in {\cal{N}}_{i}}exp(\beta \times|Pearson(\bm{W}\vec{h}_{i}, \bm{W}\vec{h}_{k})|)}
    \label{eq:Pearson atten}
\end{equation}
where $\beta \in \mathbb{R}$ is a trainable scaling factor~\cite{thekumparampil2018attention}, $Pearson(x, y) = \frac{Cov(x, y)}{\sigma_x\sigma_y}$, $Cov$ is the covariance, and $\sigma$ is the standard deviation. Following the attention coefficients normalization, node aggregation is then performed on the edges with exactly the direct neighbors of the central node and the central node itself:
\begin{equation} 
    \vec{h'}_{i} = \sum_{j \in {\cal{N}}_{i}}\alpha_{ij}\bm{W}\vec{h}_{j}
    \label{eq:central node feature}
\end{equation} 
where ${\cal{N}}_{i}$ is the neighborhood of node $i$ in the graph. 

We then use the multi-head attention, as recommended by~\cite{zheng2020gman}, which is beneficial in stabilizing the self-attention learning process. Afterward, two dense layers with a softmax are 
employed to recognize sleep stages.

\subsection{PearNet Loss Function}
In various sleep datasets, we observe that sleep stages are imbalanced, and a standard multi-class cross-entropy loss shows a bias toward the majority classes due to equally penalizing miss-classifications. To address the above issue, we apply a class-aware loss function using weighted cross-entropy~\cite{eldele2021attention} as follows:
\begin{equation}
    \mathcal{L}_{cross\_entropy} = -\frac{1}{M}\sum_{k=1}^{K}\sum_{i=1}^{I}\omega_{k}y_{i}^{k}log(\hat{y}_{i}^{k}) 
    \label{eq:cross-entropy loss}
\end{equation}
where $y_{i}^{k}$ is the ground truth of $i$-th sample for the class $k$, $\hat{y}_{i}^{k}$ is the prediction of $i$-th sample for the class $k$, $K$ is the number of classes, $M$ is the total number of samples, and $\omega_{k}$ represents the weight assigned to the class $k$. Finally, the PearNet loss function for the entire model is defined as:
\begin{equation}
    \mathcal{L}_{loss} =  \mathcal{L}_{vif}+ \mathcal{L}_{cross\_entropy}
\end{equation}

\subsection{Discussion}
\subsubsection{Adjacency matrix learning:} 
A number of previous studies have used Pearson correlations to derive the adjacency matrix~\cite{sun2014disrupted,tijms2014single,xue2019towards} through setting a threshold value. However, the Pearson correlation implies only the positive and negative correlation between graph nodes but not the internal connection relationship~\cite{neufeld2005whether}, and thus directly using Pearson correlation to determine internal relationships of signal segments is not appropriate. 
As a result, we design adjustment factors $\bm{W_2}(\sigma(\bm{W_1}(e_{ij})))$ in Equation~\eqref{eq:adjacency matrix} to adjust the Pearson correlation so that segments' internal relationships can be learned.

\subsubsection{Graph attention coefficient:}
By measuring the cosine similarity, the study~\cite{thekumparampil2018attention} captures how $j$ is relevant to $i$ as follows:
\begin{equation} 
    \alpha_{ij} = \frac{exp(\beta \times\cos(\bm{W}\vec{h}_{i}, \bm{W}\vec{h}_{j}))}{\sum_{k \in {\cal{N}}_{i}}exp(\beta \times\cos(\bm{W}\vec{h}_{i}, \bm{W}\vec{h}_{k}))}
    \label{eq:Cosin atten}
\end{equation}
where $\beta \in \mathbb{R}$ is a trainable scaling factor, and $\cos$ is the cosine similarity.
In the study~\cite{velivckovic2017graph}, the attention coefficient is learned via a single-layer feed-forward neural network, which can be expressed as:
\begin{small}
\begin{equation} 
    \alpha_{ij} = \frac{exp(LeakyReLU(\vec{v}^{T}[\bm{W}\vec{h}_{i}|| \bm{W}\vec{h}_{j}]))}{\sum_{k \in {\cal{N}}_{i}}exp(LeakyReLU(\vec{v}^{T}[\bm{W}\vec{h}_{i}|| \bm{W}\vec{h}_{k}]))}
    \label{eq:GAT atten}
\end{equation}
\end{small}
where $\vec{v}^{T}$ is the weight vector of single-layer feed-forward neural network and $||$ is the concatenation operation. Considering our proposed node generation module is constructed by VIF derived from Pearson correlation, as well as the fact that positive and negative correlations are important when learning how nodes are connected, we apply the Pearson similarity to compute the attention coefficient.

\section{Experiments}
\subsection{Datasets and Experimental Setting}
In this study, two publicly available datasets, Sleep-EDF-20 and Sleep-EDF-78~\cite{goldberger2000physiobank}, are used. The former consists of data files for 20 subjects, while the latter includes 78 subjects. Specifically, each PSG file contains two EEG channels (Fpz-Cz, Pz-Oz) with 100 Hz sampling frequencies: one EOG channel, and one chin EMG channel. In order to construct and evaluate the internal relationship between the segments of an individual electrode signal in a specific brain region, we use the data with the Fpz-Cz channel as the input for our experiments following the previous studies~\cite{supratak2017deepsleepnet,mousavi2019sleepeegnet,sun2018deep,phan2018joint}.

The hyperparameters of the experiment are shown in Table~\ref{tab:Experiment hyperparameter setting}. We implement PearNet using Pytorch 3.7 under Ubuntu 20.04 with two Nvidia GeForce RTX 2080Ti GPUs.

\subsection{Baseline Methods}
We compare our model with five baselines:
\begin{itemize}
    \item \textbf{DeepSleepNet~\cite{supratak2017deepsleepnet}:} The model exploits a custom CNN architecture followed by an LSTM with a residual connection for sleep stage classification.
    \item \textbf{ResnetLSTM~\cite{sun2018deep}:} The model implements a ResNet architecture for feature extraction, followed by an LSTM to classify EEG signals into different sleep stages.
    \item \textbf{MultitaskCNN~\cite{phan2018joint}:} The model starts by converting the raw EEG signals into power spectrum images, and then applies a joint classification and prediction technique using a multi-task CNN architecture for identifying sleep stages.
    \item \textbf{SleepEEGNet~\cite{mousavi2019sleepeegnet}:} This model employs the same CNN architecture as DeepSleepNe~\cite{supratak2017deepsleepnet} followed by an encoder-decoder with attention mechanism.
    \item \textbf{AttnSleepNet~\cite{eldele2021attention}:} The model starts with a feature extraction module that draws on a multi-resolution convolutional neural network and adaptive feature recalibration, followed by a temporal context encoder that uses a multi-head attention mechanism to capture the temporal dependency feature for recognizing sleep stages.
\end{itemize}

\begin{table}[ht]
	\centering
	\caption{Experimental hyperparameter setting}
	\scalebox{1.0}{	
	\begin{tabular}{ll}
		\toprule
		\toprule
		Hyperparameter description & Values \cr
		\cmidrule(lr){1-2}
		Number of the base segments & 5    \cr 
		Level of the base segment expansions & 2   \cr
		Number of Multi-Head Self-Attention & 3   \cr
		Number of training epochs & 100    \cr
		Batch size & 120    \cr
		K-fold cross validation & 20    \cr
		Dropout probability & 0.5    \cr
		Learning rate & 0.001    \cr
		Optimizer & AdamW with amsgrad  \cr
		\bottomrule
		\bottomrule
	\end{tabular}}
	\label{tab:Experiment hyperparameter setting}
\end{table}

\begin{table*}[htp]
	\centering
	\caption{The performance comparison of the state-of-the-art approaches}
    \scalebox{1.1}{
	\begin{tabular}{clcccccccc}
		\toprule
		\toprule
		\multirow{2}{*}{Datasets}&
		\multirow{2}{*}{Method}&
		\multicolumn{2}{c}{Overall results}& 
		\multicolumn{5}{c}{F1-score for each class} \cr
		\cmidrule(lr){3-9}
		& & Accuracy & MF1 & Wake & N1 & N2 & N3 & REM \cr 
		\cmidrule(lr){1-9}
		\multirow{6}{*}{Sleep-EDF-20} 
		&DeepSleepNet~\cite{supratak2017deepsleepnet} & 81.9 & 76.6 & 86.7 & 45.5 & 85.1 & 83.3 & 82.6   \cr
		&ResnetLSTM~\cite{sun2018deep} & 82.5 & 73.7 & 86.5 & 28.4 & 87.7 & 89.8 & 76.2   \cr
		&MultitaskCNN~\cite{phan2018joint} & 83.1 & 75.0 & 87.9 & 33.5 & 87.5 & 85.5 & 80.3   \cr
		&SleepEEGNet~\cite{mousavi2019sleepeegnet} & 81.5 & 76.6 & 89.4 & 44.4 & 84.7 & 84.6 & 79.6    \cr
		&AttnSleepNet~\cite{eldele2021attention} & 84.4 & 78.1 & 89.7 & 42.6 & 88.8 & 90.2 & 79.0   \cr
		&PearNet (\textit{ours}) & \textbf{84.9} & \textbf{79.3} & \textbf{90.0} & \textbf{46.2} & \textbf{89.1} & \textbf{90.3} & \textbf{80.6}   \cr
		\cmidrule(lr){1-9}
		\multirow{6}{*}{Sleep-EDF-78} 
		&DeepSleepNet~\cite{supratak2017deepsleepnet} & 77.8 & 71.8 & 90.9 & 45.0 & 79.2 & 72.7 & 71.1   \cr
		&ResnetLSTM~\cite{sun2018deep} & 78.9 & 71.4 & 90.7 & 34.7 & 83.6 & 80.9 & 67.0   \cr
		&MultitaskCNN~\cite{phan2018joint} & 79.6 & 72.8 & 90.9 & 39.7 & 83.2 & 76.6 & 73.5   \cr
		&SleepEEGNet~\cite{mousavi2019sleepeegnet} & 74.2 & 69.6 & 89.8 & 42.1 & 75.2 & 70.4 & 70.6    \cr
		&AttnSleepNet~\cite{eldele2021attention} & 81.3 & 75.1 & 92.0 & 42.0 & \textbf{85.0} & 82.1 & \textbf{74.2}   \cr
		&PearNet (\textit{ours}) & \textbf{81.6} & \textbf{75.3} & \textbf{92.1} & \textbf{43.0} & 84.9 & \textbf{82.6} & 74.0   \cr
		\bottomrule
		\bottomrule
	\end{tabular}}
	\label{tab:SOTA performance}
\end{table*}

\subsection{Comparison and Result Analysis}
Table~\ref{tab:SOTA performance} shows the comparison of PearNet with five baseline algorithms for recognising sleep stages on the Sleep-EDF-20 and the Sleep-EDF-78. Specifically, PearNet's accuracy and macro F1-score (MF1) are $84.9$ and $79.3$ on Sleep-EDF-20, and $81.4$ and $75.3$ on Sleep-EDF-78. From the results, it proves that the adaptively learned graph representation for the underlying EEG signals is more suitable than the grid representation as used by CNN and LSTM~\cite{supratak2017deepsleepnet,mousavi2019sleepeegnet,sun2018deep,phan2018joint,eldele2021attention}. Further, PearNet's MF1 for each class performs better than all baseline methods on Sleep-EDF-20, while for Sleep-EDF-78, the results of Wake, $N1$, and $N3$ are the best. Due to the possibly ineffective settings of  the number of the base segments and the level of the hierarchical feature extraction for Sleep-EDF-78, PearNet's MF1 for $N2$ ($84.9$) and $REM$ ($74$) are lower than those of AttnSleeNet ($85$ and $74.2$, respectively). Moreover, while 
PearNet is the best over all baseline methods for class $N1$, the MF1 is still below 50, indicating that it is worth considering both internal signal segment relationships and external connections across different brain regions to recognize sleep stages, which is our next-stage work.

\subsection{Ablation and Sensitivity Experiment}
\begin{table}[htp]
	\centering
    \caption{Ablation and Sensitivity Study}
    \scalebox{1}{
	\begin{tabular}{lcc}
		\toprule
		\toprule
		\multirow{2}{*}{Method}&
		\multicolumn{2}{c}{Overall results} \cr
		\cmidrule(lr){2-3}
		& Accuracy & MF1 \cr 
		\cmidrule(lr){1-3}
		PearNet-Base(2, 2) & 84.2 & 78.1   \cr
		PearNet-Base(5, 2) & \textbf{84.9} & \textbf{79.3}  \cr
		PearNet-Base(8, 2) & 84.4 & 78.5  \cr
		\cmidrule(lr){1-3}
		PearNet-Level(5, 0) & 84.6 & 78.3  \cr
		PearNet-Level(5, 2) & \textbf{84.9} & \textbf{79.3}  \cr
		PearNet-Level(5, 3) & 84.8 & 78.7  \cr
		\cmidrule(lr){1-3}
		PearNet-Atten(5, 2) with GAT attention & 84.4 & 78.5   \cr   
		PearNet-Atten(5, 2) with AGNN attention & 84.2 & 78.3  \cr
		PearNet-Atten(5, 2) with Pearson attention & \textbf{84.9} & \textbf{79.3} \cr
		\cmidrule(lr){1-3}
		PearNet-VIF(5, 2) without VIF Loss & 84.7 & 78.7  \cr
		PearNet-VIF(5, 2) with VIF Los & \textbf{84.9} & \textbf{79.3}  \cr
		\bottomrule
		\bottomrule
	\end{tabular}
	}
	\label{tab:Ablation Experiment}
\end{table}

To further investigate the effectiveness of PearNet, we conduct an ablation and sensitivity study on the Sleep-EDF-20 dataset as shown in Table~\ref{tab:Ablation Experiment}. Specifically, we derive a set of model variants from PearNet as follows:
\begin{itemize}
    \item [1)] 
    \textbf{PearNet-Base($x$ base segments, 2 levels of the feature extraction):} PearNet-Base($x$, 2) is investigated in terms of the input nodes that are generated from $x$ base segments with 2 levels of feature extraction. We vary
    $x$ from $\{2, 5, 8\}$.
    \item [2)] 
    \textbf{PearNet-Level(5 base segments, $x$ levels of the feature extraction):} PearNet-Level(5, $x$) is investigated in terms of the input nodes that are generated from 5 base segments with $x$ levels of feature extraction. We vary
    $x$ from $\{0, 2, 3\}$.
    \item [3)] 
    \textbf{PearNet-Atten(5, 2) with $x$ attention:} PearNet-Atten(5, 2) with the input nodes that are generated from 5 basis segments and 2 levels of feature extraction is investigated in terms of the $x$ types of attention mechanisms. We vary $x$ from $\{$GAT~attention~\cite{velivckovic2017graph}, AGNN~attention~\cite{thekumparampil2018attention}, Pearson~attention$\}$ 
    \item [4)] 
    \textbf{PearNet-VIF(5, 2) with (without) VIF Loss:} PearNet-VIF(5, 2) with the input nodes that are generated from 5 basis segments and 2 levels of feature extraction is investigated with and without the VIF loss.
\end{itemize}

PearNet-Base($x$, 2) evaluates PearNet's performance with various numbers of base segments under the same levels of feature extraction. 
The results verify that PearNet can
recognize the sleep stages under various base segments; meanwhile, we notice that PearNet with 5 and 8 base segments performs better than PearNet with 2 segments. PearNet with 8 base segments, however, has a lower accuracy and MF1 ($84.4$ and $78.5$) than PearNet with 5 base segments ($84.9$ and $79.3$). The findings imply that, with a fixed window size of 30 seconds, more segments with fewer data points 
are not always better, since important signal patterns can be missed.

PearNet-Level(5, $x$) is evaluated at different levels of feature extraction with a fixed number of base segments. Since higher level nodes have better temporal resolution, PearNet with 2-level feature extraction has a higher accuracy and MF1 ($84.9$ and $79.3$) than PearNet with 0-level feature extraction. Because the proposed hierarchical feature extraction method uses zero padding to keep successive layers the same length, the less number of segments from the lower layer generates a greater number of zero-padding related nodes. It causes more redundant information, negatively affecting recognition accuracy. We can observe that PearNet with 3-level feature extraction and 5 base segments has a lower accuracy ($84.8$) and MF1 ($78.7$) than PearNet with 2-level feature extraction and 5 base segments ($84.9$ and $79.3$).

PearNet-Atten(5, 2) with $x$ attention evaluates the performance of PearNet with different attention mechanisms. Pearson attention achieves the best performance ($84.9$ on accuracy and $79.3$ on MF1), which is consistent with our design intent. Pearson correlation is used to train the key modules of PearNet (node generation with VIF loss and graph structure). It may lead to better performance after an end-to-end training using a Pearson correlation related attention.

PearNet-VIF(5, 2) with (resp. without) VIF Loss examines PearNet's performance with (resp. without) the VIF loss under the original setting. 
The results show that PearNet improves its performance once VIF loss is adopted, 
and optimizing the multicollinearity problem reduces the detrimental effects of highly correlated nodes on model performance.

From these findings, we can conclude that using Pearson correlation for node generation, edge construction, and graph attention is suitable for graph representation of the underlying data. As a promising future direction, our proposed PearNet architecture can be adopted to study its suitability for time-series applications in general including activity recognition, driving fatigue detection, and many other fields.

\section{Conclusion}
In this paper, we propose a novel deep graph attention network called PearNet for identifying sleep stages. The main advantage of PearNet is that it models the internal spatial-temporal relationships between the segments of individual electrode signals in a specific brain region. Specifically, PearNet 
is capable of generating graph nodes from the signal segments and adaptively learning the graph structure to determine the pairwise connections between segments. Experimental results demonstrate that PearNet achieves the state-of-the-art performance. PearNet offers an insight into analyzing the signals with regard to the internal relationships of the EEG signal and it is worth investigating how to apply Pearson correlation-based graph networks to other time series data.

\section{Acknowledgements}
This work is in part supported by an Australian Research Council (ARC) Discovery Project (DP210102447), an ARC Linkage Project (LP190100676), and a DATA61 project (Data61 CRP C020996).

\bibliography{ref.bib}

\begin{thebibliography}{10}
\providecommand{\url}[1]{#1}
\csname url@samestyle\endcsname
\providecommand{\newblock}{\relax}
\providecommand{\bibinfo}[2]{#2}
\providecommand{\BIBentrySTDinterwordspacing}{\spaceskip=0pt\relax}
\providecommand{\BIBentryALTinterwordstretchfactor}{4}
\providecommand{\BIBentryALTinterwordspacing}{\spaceskip=\fontdimen2\font plus
\BIBentryALTinterwordstretchfactor\fontdimen3\font minus
  \fontdimen4\font\relax}
\providecommand{\BIBforeignlanguage}[2]{{%
\expandafter\ifx\csname l@#1\endcsname\relax
\typeout{** WARNING: IEEEtran.bst: No hyphenation pattern has been}%
\typeout{** loaded for the language `#1'. Using the pattern for}%
\typeout{** the default language instead.}%
\else
\language=\csname l@#1\endcsname
\fi
#2}}
\providecommand{\BIBdecl}{\relax}
\BIBdecl

\bibitem{fallmann2019computational}
S.~Fallmann and L.~Chen, ``Computational sleep behavior analysis: A survey,''
  \emph{IEEE Access}, vol.~7, pp. 142\,421--142\,440, 2019.

\bibitem{roebuck2013review}
A.~Roebuck, V.~Monasterio, E.~Gederi, M.~Osipov, J.~Behar, A.~Malhotra,
  T.~Penzel, and G.~Clifford, ``A review of signals used in sleep analysis,''
  \emph{Physiological Measurement}, vol.~35, no.~1, pp. R1--R57, 2014.

\bibitem{moser2009sleep}
D.~Moser, P.~Anderer, G.~Gruber, S.~Parapatics, E.~Loretz, M.~Boeck,
  G.~Kloesch, E.~Heller, A.~Schmidt, H.~Danker-Hopfe \emph{et~al.}, ``Sleep
  classification according to aasm and rechtschaffen \& kales: effects on sleep
  scoring parameters,'' \emph{Sleep}, vol.~32, no.~2, pp. 139--149, 2009.

\bibitem{jia2020graphsleepnet}
Z.~Jia, Y.~Lin, J.~Wang, R.~Zhou, X.~Ning, Y.~He, and Y.~Zhao, ``Graphsleepnet:
  Adaptive spatial-temporal graph convolutional networks for sleep stage
  classification.'' in \emph{Proceedings of the International Joint Conference
  on Artificial Intelligence}, 2020, pp. 1324--1330.

\bibitem{lajnef2015learning}
T.~Lajnef, S.~Chaibi, P.~Ruby, P.-E. Aguera, J.-B. Eichenlaub, M.~Samet,
  A.~Kachouri, and K.~Jerbi, ``Learning machines and sleeping brains: automatic
  sleep stage classification using decision-tree multi-class support vector
  machines,'' \emph{Journal of Neuroscience Methods}, vol. 250, pp. 94--105,
  2015.

\bibitem{fraiwan2012automated}
L.~Fraiwan, K.~Lweesy, N.~Khasawneh, H.~Wenz, and H.~Dickhaus, ``Automated
  sleep stage identification system based on time--frequency analysis of a
  single eeg channel and random forest classifier,'' \emph{Computer Methods and
  Programs in Biomedicine}, vol. 108, no.~1, pp. 10--19, 2012.

\bibitem{gunecs2010efficient}
S.~G{\"u}ne{\c{s}}, K.~Polat, and {\c{S}}.~Yosunkaya, ``Efficient sleep stage
  recognition system based on eeg signal using k-means clustering based feature
  weighting,'' \emph{Expert Systems with Applications}, vol.~37, no.~12, pp.
  7922--7928, 2010.

\bibitem{phan2018joint}
H.~Phan, F.~Andreotti, N.~Cooray, O.~Y. Ch{\'e}n, and M.~De~Vos, ``Joint
  classification and prediction cnn framework for automatic sleep stage
  classification,'' \emph{IEEE Transactions on Biomedical Engineering},
  vol.~66, no.~5, pp. 1285--1296, 2018.

\bibitem{phan2019seqsleepnet}
H.~Phan, F.~Andreotti, N.~Cooray, O.~Y. Chén, and M.~De~Vos, ``Seqsleepnet:
  End-to-end hierarchical recurrent neural network for sequence-to-sequence
  automatic sleep staging,'' \emph{IEEE Transactions on Neural Systems and
  Rehabilitation Engineering}, vol.~27, no.~3, pp. 400--410, 2019.

\bibitem{sun2018deep}
Y.~Sun, B.~Wang, J.~Jin, and X.~Wang, ``Deep convolutional network method for
  automatic sleep stage classification based on neurophysiological signals,''
  in \emph{Proceedings of the International Congress on Image and Signal
  Processing, BioMedical Engineering and Informatics}, 2018, pp. 1--5.

\bibitem{supratak2017deepsleepnet}
A.~Supratak, H.~Dong, C.~Wu, and Y.~Guo, ``Deepsleepnet: A model for automatic
  sleep stage scoring based on raw single-channel eeg,'' \emph{IEEE
  Transactions on Neural Systems and Rehabilitation Engineering}, vol.~25,
  no.~11, pp. 1998--2008, 2017.

\bibitem{mousavi2019sleepeegnet}
S.~Mousavi, F.~Afghah, and U.~R. Acharya, ``{SleepEEGNet: Automated sleep stage
  scoring with sequence to sequence deep learning approach},'' \emph{PloS One},
  vol.~14, no.~5, p. e0216456, 2019.

\bibitem{eldele2021attention}
E.~Eldele, Z.~Chen, C.~Liu, M.~Wu, C.-K. Kwoh, X.~Li, and C.~Guan, ``An
  attention-based deep learning approach for sleep stage classification with
  single-channel eeg,'' \emph{IEEE Transactions on Neural Systems and
  Rehabilitation Engineering}, vol.~29, pp. 809--818, 2021.

\bibitem{eldele2021time}
E.~Eldele, M.~Ragab, Z.~Chen, M.~Wu, C.~K. Kwoh, X.~Li, and C.~Guan,
  ``Time-series representation learning via temporal and contextual
  contrasting,'' in \emph{Proceedings of the International Joint Conference on
  Artificial Intelligence}, 2021, pp. 2352--2359.

\bibitem{jia2021salientsleepnet}
Z.~Jia, Y.~Lin, J.~Wang, X.~Wang, P.~Xie, and Y.~Zhang, ``Salientsleepnet:
  Multimodal salient wave detection network for sleep staging,'' \emph{arXiv
  preprint arXiv:2105.13864}, 2021.

\bibitem{velivckovic2017graph}
\BIBentryALTinterwordspacing
P.~Veli{\v{c}}kovi{\'{c}}, G.~Cucurull, A.~Casanova, A.~Romero, P.~Li{\`{o}},
  and Y.~Bengio, ``{Graph Attention Networks},'' in \emph{Proceedings of the
  International Conference on Learning Representations}, 2018. [Online].
  Available: \url{https://openreview.net/forum?id=rJXMpikCZ}
\BIBentrySTDinterwordspacing

\bibitem{hamilton2017inductive}
W.~L. Hamilton, R.~Ying, and J.~Leskovec, ``Inductive representation learning
  on large graphs,'' in \emph{Proceedings of the International Conference on
  Neural Information Processing Systems}, 2017, pp. 1025--1035.

\bibitem{olbrich2011multiple}
E.~Olbrich, J.~C. Claussen, and P.~Achermann, ``The multiple time scales of
  sleep dynamics as a challenge for modelling the sleeping brain,''
  \emph{Philosophical Transactions of the Royal Society A: Mathematical,
  Physical and Engineering Sciences}, vol. 369, no. 1952, pp. 3884--3901, 2011.

\bibitem{yap2017oscillatory}
M.~H. Yap, M.~J. Grabowska, C.~Rohrscheib, R.~Jeans, M.~Troup, A.~C. Paulk,
  B.~Van~Alphen, P.~J. Shaw, and B.~Van~Swinderen, ``Oscillatory brain activity
  in spontaneous and induced sleep stages in flies,'' \emph{Nature
  Communications}, vol.~8, no.~1, pp. 1--15, 2017.

\bibitem{hu2018squeeze}
J.~Hu, L.~Shen, and G.~Sun, ``Squeeze-and-excitation networks,'' in
  \emph{Proceedings of the IEEE Conference on Computer Vision and Pattern
  Recognition}, 2018, pp. 7132--7141.

\bibitem{gu2018stack}
J.~Gu, J.~Cai, G.~Wang, and T.~Chen, ``Stack-captioning: coarse-to-fine
  learning for image captioning,'' in \emph{Proceedings of the AAAI Conference
  on Artificial Intelligence}, 2018, pp. 6837--6844.

\bibitem{bai2018empirical}
S.~Bai, J.~Z. Kolter, and V.~Koltun, ``An empirical evaluation of generic
  convolutional and recurrent networks for sequence modeling,'' \emph{arXiv
  preprint arXiv:1803.01271}, 2018.

\bibitem{shen2020stable}
Z.~Shen, P.~Cui, T.~Zhang, and K.~Kunag, ``Stable learning via sample
  reweighting,'' in \emph{Proceedings of the AAAI Conference on Artificial
  Intelligence}, 2020, pp. 5692--5699.

\bibitem{rui2020geometry}
Z.~Rui, G.~Zongyuan, D.~Simon, S.~Sridha, and F.~Clinton,
  ``Geometry-constrained car recognition using a 3d perspective network,'' in
  \emph{Proceedings of the AAAI Conference on Artificial Intelligence}, 2020,
  pp. 1161--1168.

\bibitem{stine1995graphical}
R.~A. Stine, ``Graphical interpretation of variance inflation factors,''
  \emph{The American Statistician}, vol.~49, no.~1, pp. 53--56, 1995.

\bibitem{jiang2013graph}
B.~Jiang, C.~Ding, B.~Luo, and J.~Tang, ``Graph-laplacian pca: Closed-form
  solution and robustness,'' in \emph{Proceedings of the IEEE Conference on
  Computer Vision and Pattern Recognition}, 2013, pp. 3492--3498.

\bibitem{thekumparampil2018attention}
K.~K. Thekumparampil, C.~Wang, S.~Oh, and L.-J. Li, ``Attention-based graph
  neural network for semi-supervised learning,'' \emph{arXiv preprint
  arXiv:1803.03735}, 2018.

\bibitem{zheng2020gman}
C.~Zheng, X.~Fan, C.~Wang, and J.~Qi, ``Gman: A graph multi-attention network
  for traffic prediction,'' in \emph{Proceedings of the AAAI Conference on
  Artificial Intelligence}, 2020, pp. 1234--1241.

\bibitem{sun2014disrupted}
Y.~Sun, Q.~Yin, R.~Fang, X.~Yan, Y.~Wang, A.~Bezerianos, H.~Tang, F.~Miao, and
  J.~Sun, ``Disrupted functional brain connectivity and its association to
  structural connectivity in amnestic mild cognitive impairment and
  alzheimer’s disease,'' \emph{PloS one}, vol.~9, no.~5, p. e96505, 2014.

\bibitem{tijms2014single}
B.~M. Tijms, H.~M. Yeung, S.~A. Sikkes, C.~M{\"o}ller, L.~L. Smits, C.~J. Stam,
  P.~Scheltens, W.~M. van~der Flier, and F.~Barkhof, ``Single-subject gray
  matter graph properties and their relationship with cognitive impairment in
  early-and late-onset alzheimer's disease,'' \emph{Brain Connectivity},
  vol.~4, no.~5, pp. 337--346, 2014.

\bibitem{xue2019towards}
H.~Xue, J.~Peng, and X.~Shang, ``Towards gene function prediction via
  multi-networks representation learning,'' in \emph{Proceedings of the AAAI
  Conference on Artificial Intelligence}, 2019, pp. 10\,069--10\,070.

\bibitem{neufeld2005whether}
E.~Neufeld and S.~Kristtorn, ``Whether non-correlation implies non-causation.''
  in \emph{Proceedings of the Florida Artificial Intelligence Research Society
  Conference}, 2005, pp. 772--777.

\bibitem{goldberger2000physiobank}
A.~L. Goldberger, L.~A. Amaral, L.~Glass, J.~M. Hausdorff, P.~C. Ivanov, R.~G.
  Mark, J.~E. Mietus, G.~B. Moody, C.-K. Peng, and H.~E. Stanley, ``Physiobank,
  physiotoolkit, and physionet: components of a new research resource for
  complex physiologic signals,'' \emph{circulation}, vol. 101, no.~23, pp.
  e215--e220, 2000.

\end{thebibliography}
\bibliographystyle{IEEEtran}

\end{document}